# Association of macroscopic laboratory testing and micromechanics modelling for the evaluation of the poroelastic parameters of a hardened cement paste

Siavash Ghabezloo[*]

*Université Paris-Est, UR Navier, CERMES, Ecole des Ponts ParisTech, Marne la Vallée, France*

## Abstract

The results of a macro-scale experimental study performed on a hardened class G cement paste [Ghabezloo et al. (2008) Cem. Con. Res. (38) 1424-1437] are used in association with the micromechanics modelling and homogenization technique for evaluation of the complete set of poroelastic parameters of the material. The experimental study consisted in drained, undrained and unjacketed isotropic compression tests. Analysis of the experimental results revealed that the active porosity of the studied cement paste is smaller than its total porosity. A multi-scale homogenization model, calibrated on the experimental results, is used to extrapolate the poroelastic parameters to cement pastes prepared with different water-to-cement ratio. The notion of cement paste active porosity is discussed and the poroelastic parameters of hardened cement paste for an ideal, perfectly drained condition are evaluated using the homogenization model.

---

[*] Siavash Ghabezloo, CERMES, Ecole des Ponts ParisTech, 6-8 avenue Blaise Pascal, Cité Descartes, 77455 Champs-sur-Marne, Marne la Vallée cedex 2, France. Email: siavash.ghabezloo@enpc.fr

# 1. Introduction

Hardened cement paste is a complex and evolving porous material resulting from chemical reactions, referred to as hydration, between cement clinker and water. The physical and mechanical properties of the hardened cement paste depend upon clinker composition, water-to-cement ratio, cement age and curing conditions. The microstructure of this material is composed of several solid phases and of a porous phase which contains some quantity of adsorbed water. The size of pores in the microstructure of the hardened cement paste covers an impressive range from nanometric gel pores to micrometric capillary pores and even millimetric air voids. Due to the complexity of this microstructure, the description of the behaviour of cementitious materials within the framework of the classical theory of poromechanics is considered as an open question, as mentioned by Ulm *et al*. [1]. Recent progress in advanced micromechanical testing methods, such as nano-indentation tests, have provided direct measurements of the elastic properties of the different phases of the microstructure and has made it possible to estimate the poroelastic properties of the hardened cement paste using the theory of micro-poro-mechanics and homogenization methods [1][2][3]. These evaluations of the poroelastic properties are mostly supported by the experimental results obtained using non-destructive methods such as, among others, elastic resonance and ultrasonic wave velocity measurements, but have been rarely validated against mechanical loading tests results. Indeed very few experimental results of classical poromechanics tests, such as drained, undrained and unjacketed compression tests, can be found in the literature for the hardened cement paste. Recently, Ghabezloo *et al*. [4][5][6][7] studied experimentally the thermo-poro-mechanical behaviour of a hardened cement paste by performing these classical poromechanics tests, as well as drained and undrained heating tests and permeability evaluation tests. As an answer to the question raised by Ulm *et al*. [1], the results presented in Ghabezloo et al. [4] show clearly that the hardened cement paste can be indeed considered as a poromechanics material. An important difficulty encountered when testing hardened cement paste is its very low permeability which makes the poromechanical tests long and expensive. The experimental tests presented by Ghabezloo *et al*. [4][5][6][7] have been performed on a particular cement paste. This cement paste is prepared with class G cement at a water-to-cement ratio equal to 0.44 that was hydrated at 90°C for at least 90 days in saturated condition. It is well known that the physical and mechanical properties of hardened cement paste vary with clinker composition, water-to-cement ratio, cement age and curing conditions. Consequently the complete characterization of the poromechanical properties of this material in an experimental study would be extremely expensive and time consuming. The characterization of the poroelastic properties of the particular hardened cement paste studied by Ghabezloo *et al*. [4][5][6][7] can be extrapolated to evaluate these properties for various water-to-cement ratios by means of micromechanics modelling and homogenization techniques and this is the aim of the present paper.

The paper is organized in seven sections. After this introduction, the second section presents briefly the theoretical framework of Macro- and Micro-poroelasticity and the homogenization method. A summary of the results of the macro-scale experimental study is presented in the third section. The fourth section deals with the microstructure of cement paste and evaluation of volume fractions of its different phases. The homogenization of the poroelastic properties of the cement paste is presented in the fifth section. First the model is calibrated on the basis of the results of the macro-scale experimental study, and then the homogenization model is used to extrapolate these experimental results for the cement pastes with different water-to-cement ratios. A discussion is presented in the sixth section on the active porosity of the cement paste from the poromechanics point of view. The last section is dedicated to the concluding remarks.

## 2. Theoretical framework

This paper associates the results of a macro-scale experimental study with the tools of micro-poroelasticity theory and the homogenization technique in order to characterize the complete set of poroelastic parameters of a cement paste. It is thus necessary to present the theoretical framework used in the macro-scale experimental study, the one used for the micromechanics modelling and homogenization technique, as well as the link between the parameters in these two scales.

### *2.1. Macro-poroelasticity*

We present here the framework used to describe the macroscopic elastic volumetric behaviour of a porous material which is heterogeneous at the micro-scale. The theoretical basis of the formulation has been presented in the milestone papers and textbooks of Biot and Willis [8], Brown and Korringa [9], Rice and Cleary [10], Zimmerman [11], Berryman [12], Detournay and Cheng [13], Vardoulakis and Sulem [14], Coussy [15]. This framework is briefly recalled here to clarify the mathematical and physical significance of the poromechanical parameters evaluated in our experimental program. A more detailed presentation of this framework is given in Ghabezloo *et al.* [4][6] and Ghabezloo [7].

A fluid saturated porous material can be seen as a mixture of a solid phase and a fluid phase. The solid phase may be itself made up of several constituents. The Lagrangian porosity $\phi$ is defined as the pore volume $V_\phi$ per unit volume of porous material in the reference state $V_0$ (also called pore volume fraction), $\phi = V_\phi / V_0$. For a saturated sample under an isotropic state of stress $\Sigma = 1/3 \boldsymbol{\Sigma} : \mathbf{1}$ (positive in compression), we choose two independent variables for characterizing the volumetric behaviour: the pore pressure $p$ and the differential pressure $\Sigma_d$ which is equivalent to Terzaghi effective stress $\Sigma_d = \Sigma - p$. The expression of the variations of the total volume $V$ and of the pore volume $V_\phi$ introduces four parameters:

$$-\frac{dV}{V_0} = \frac{1}{K_d} d\Sigma_d + \frac{1}{K_s} dp \qquad (1)$$

$$-\frac{dV_\phi}{V_{\phi 0}} = \frac{1}{K_p} d\Sigma_d + \frac{1}{K_\phi} dp \qquad (2)$$

$$\frac{1}{K_d} = -\frac{1}{V_0}\left(\frac{\partial V}{\partial \Sigma_d}\right)_p \quad , \quad \frac{1}{K_p} = -\frac{1}{V_{\phi 0}}\left(\frac{\partial V_\phi}{\partial \Sigma_d}\right)_p \qquad (3)$$

$$\frac{1}{K_s} = -\frac{1}{V_0}\left(\frac{\partial V}{\partial p}\right)_{\Sigma_d} \quad , \quad \frac{1}{K_\phi} = -\frac{1}{V_{\phi 0}}\left(\frac{\partial V_\phi}{\partial p}\right)_{\Sigma_d} \qquad (4)$$

Equation (3) corresponds to a drained isotropic compression test in which the pore pressure is constant in the sample. The variations of the total volume of the sample $V$ and of the volume of the pore space $V_\phi$ with respect to the applied confining pressure give the drained bulk modulus $K_d$ and the modulus $K_p$. Equation (4) corresponds to the so-called unjacketed compression test, in which equal increments of confining pressure and pore pressure are simultaneously applied to the sample, as if the sample was submerged without a jacket, into a fluid under the pressure $p$. The differential pressure $\Sigma_d$ in this condition remains constant. Neglecting the deformation of the jacket, the change in volumetric strain with the applied pressure gives the unjacketed modulus $K_s$. The variation of the pore volume of the sample in this test, evaluated from the quantity of fluid exchanged between the sample and the pore pressure generator when applying equal increments of confining pressure and pore pressure could in principle give the modulus $K_\phi$. However experimental evaluation of this parameter is very difficult as the volume of the exchanged fluid has to be corrected for the effect of fluid compressibility and also for the effect of the deformations of the pore pressure generator and drainage system. In the case of a porous material which is homogeneous and isotropic at the micro-scale, the sample would deform in an unjacketed test as if all the pores were filled with the solid component. The skeleton and the solid component experience a uniform volumetric strain with no change of the Eulerian porosity. For such a material $K_s = K_\phi = K_m$, where $K_m$ is the bulk modulus of the single solid constituent of the porous material. In the case of a porous material that is composed of two or more solids and therefore is heterogeneous at the micro-scale, the unjacketed modulus $K_s$ is some weighted average of the bulk moduli of solid constituents [12]. The modulus $K_\phi$ for such a material has a complicated dependence on the material properties. Generally it is not bounded by the elastic moduli of the solid components and can even have a negative sign if the bulk moduli of the individual solid components are greatly different one from another [16][17].

Using Betti's reciprocal theorem one obtains the following relation between the elastic moduli [9][18]:

$$\frac{1}{K_p} = \frac{1}{\phi_0}\left(\frac{1}{K_d} - \frac{1}{K_s}\right) \qquad (5)$$

The expression of Biot effective stress coefficient $b$ can be obtained by re-writing equation (1):

$$dE = \frac{1}{K_d}(d\Sigma - bdp) \quad ; \quad b = 1 - \frac{K_d}{K_s} \qquad (6)$$

where $dE = -dV/V_0$ is the macroscopic volumetric strain increment ($E = \mathbf{E}:\mathbf{1}$). The variation of Lagrangian porosity is given by $d\phi = \phi_0 dV_\phi/V_{\phi 0}$ where the term $dV_\phi/V_{\phi 0}$ is given by equation (2). From equation (1) we have $d\Sigma_d = K_d(dE - dp/K_s)$. Replacing the term $d\Sigma_d$ in equation (2) by this expression and using equations (5) and (6) the following equation is obtained for the variation of Lagrangian porosity:

$$d\phi = -bdE + \frac{1}{N}dp \tag{7}$$

where $N$ is Biot skeleton modulus which is linking the pore pressure and porosity variations when the strains are held constant:

$$\frac{1}{N} = \left(\frac{\partial \phi}{\partial p}\right)_E = \frac{b}{K_s} - \frac{\phi_0}{K_\phi} \tag{8}$$

The fluid content $m_f$ is defined as the fluid mass per unit volume of the porous material and is given by $m_f = \phi_0 \rho_f$. Using this expression, equation (7) and knowing that $d\rho_f/\rho_f = dp/K_f$, the following relation is found for the variation of the fluid content:

$$\frac{dm_f}{\rho_f} = -bdE + \frac{1}{M}dp \quad ; \quad \frac{1}{M} = \frac{1}{N} + \frac{\phi_0}{K_f} \tag{9}$$

where $M$ is Biot modulus and $K_f$ is the fluid compression modulus. The undrained condition is defined as a condition in which the mass of the fluid phase is constant ($dm_f = 0$). In this condition, inserting $dm_f = 0$ and equation (6) in equation (9) the Skempton [19] coefficient $B$ can be identified:

$$dp = Bd\Sigma \quad ; \quad B = \left(\frac{\partial p}{\partial \Sigma}\right)_{m_f} = \frac{Mb}{K_d + Mb^2} \tag{10}$$

The expression of undrained bulk modulus $K_u$ can be found by replacing $dp$ from equation (10) in equation (6):

$$dE = \frac{1}{K_u}d\Sigma \quad ; \quad \frac{1}{K_u} = -\frac{1}{V_0}\left(\frac{\partial V}{\partial \Sigma}\right)_{m_f} = \frac{1-bB}{K_d} \tag{11}$$

Among the nine different poroelastic parameters presented in this section $K_d$, $K_s$, $K_p$, $K_\phi$, $b$, $N$, $M$, $K_u$ and $B$, evaluation of three independent parameters in addition to the porosity $\phi$ is sufficient for a complete characterization of the poroelastic behaviour of the porous material. In laboratory experiments, the three most commonly performed tests are drained, undrained and unjacketed compression tests which yield $K_d$, $K_p$, $K_u$, $B$ and $K_s$. The other poroelastic parameters can then be evaluated indirectly using relations presented in this section. On the other hand, the derivation of the equations of micro-poroelasticity and the homogenization of poroelastic parameters, which is presented in the next section, is more commonly done using $K_d$, $b$ and $N$. The presentation of the complete set of poroelastic parameters, as presented in this section, will permit to establish the link between the parameters that are easier to evaluate experimentally and the ones used more commonly in micro-poroelasticity.

## *2.2. Micro-poroelasticity and homogenization method*

The theoretical framework of the micromechanics modelling and the homogenization method used in this study has been presented Zaoui [20][21], Dormieux et al. [22][23][24], Ulm et al. [1], Château and Dormieux [25]. The aim of classical homogenization techniques is to replace an actual heterogeneous complex body by a fictitious homogeneous one that behaves globally in the same way. Continuum micromechanics is mainly concerned with statistically homogeneous materials for which it is possible to define a representative elementary volume (REV). Over the REV, the average values of local stress and strain fields in the actual heterogeneous body are equal to the macroscopic values of stress and strain fields derived by solving the boundary value problem of a homogeneous body made of this fictitious homogeneous material [21]. This requires that, for the mechanical behaviour under investigation, the characteristic length *d* of the heterogeneity and deformation mechanism to be much smaller than the size *l* of the volume element. Moreover, *l* must be sufficiently smaller than the characteristic dimension *L* of the whole body.

After the scale separation, the three steps of homogenization method as mentioned by Zaoui [21] are: *description* (or representation), *concentration* (or localization) and *homogenization* (or upscaling). The *description* step deals with identification of different "mechanical" phases of the microstructure in the REV of the heterogeneous material, and both geometrical and mechanical characteristics of these phases. A phase, in the sense of continuum micromechanics, is a material domain that can be identified at a given scale, with on-average constant material properties. The *concentration* step is concerned with the mechanical modelling of the interactions between the phases and the link between the local stress and strain fields within the REV and the macroscopic quantities of stress and strain. The last step deals with the *homogenization* of the macroscopic properties by combining the local constitutive equations, averaging the stresses and the strains over the REV and the concentration relations. Homogenization delivers the estimated values of macroscopic poroelastic properties of the REV as a function of the geometrical and mechanical properties of different phases of the microstructure of the material.

### 2.2.1. Description

The volume $V_0$ of the REV of a heterogeneous material is composed of *n* different phases with volumes $V_r$, $r = 1 \ldots n$, and volume fractions denoted by $f_r = V_r/V_0$. We consider that there is only one porous phase with volume $V_\phi$ and porosity $\phi = V_\phi/V_0$. The number of solid phases is thus $m = n - 1$ with total volume $V_s$. The tensor of elastic moduli of each phase is denoted by $\mathbf{c}_r$. In the case of isotropy of the solid phases, the tensor of elastic moduli can be written as the sum of a volumetric and a deviatoric part:

$$\mathbf{c}_r = 3k_r \mathbf{J} + 2g_r \mathbf{K} \tag{12}$$

where $k_r$ and $g_r$ are the bulk modulus and shear modulus of the phase *r* respectively. $\mathbf{J}_{ijkl} = 1/3 \, \delta_{ij}\delta_{kl}$ is the volumetric part of the fourth-order symmetric unit tensor $\mathbf{I}$ and $\mathbf{K} = \mathbf{I} - \mathbf{J}$ is the deviatoric part. $\mathbf{I}$ is defined as $\mathbf{I}_{ijkl} = 1/2\left(\delta_{ik}\delta_{jl} + \delta_{il}\delta_{jk}\right)$ and $\delta_{ij}$ stands for the Kronecker delta. The representation of the microstructure of the cement paste and the evaluation of the volume fractions of the different phases will be discussed further in section 4.2.

## 2.2.2. Concentration

The concentration problem is presented by assuming homogeneous boundary conditions on the REV (Hill [26], Hashin [27]). *Homogeneous strain boundary conditions* are associated to prescribed displacements $\underline{u}$ at the boundary of the REV as $\underline{u} = \mathbf{E} \cdot \underline{x}$, where $\underline{x}$ is the microscopic position vector and $\mathbf{E}$ is the macroscopic strain tensor. It can be shown that $\mathbf{E}$ is equal to the volume average of the microscopic compatible (i.e., derived from a displacement field) strain field $\boldsymbol{\varepsilon}(\underline{x})$ in the REV [21], $\mathbf{E} = \langle \boldsymbol{\varepsilon} \rangle_V$ where $\langle z \rangle_V = (1/V) \int_V z(\underline{x}) \mathrm{d}V$ stands for the volume average of quantity $z$ over domain $V$.

In the framework of linear elasticity, the local strain field $\boldsymbol{\varepsilon}(\underline{x})$ is related to macroscopic strain $\mathbf{E}$ through fourth-order localization tensor $\mathbf{A}(\underline{x})$:

$$\boldsymbol{\varepsilon}(\underline{x}) = \mathbf{A}(\underline{x}) : \mathbf{E} \tag{13}$$

By inserting equation (13) in the equality $\mathbf{E} = \langle \boldsymbol{\varepsilon} \rangle_V$ one obtains $\langle \mathbf{A} \rangle_V = \mathbf{I}$. For a heterogeneous material composed of homogeneous phases, a linear phase strain localization tensor can be introduced [1]:

$$\langle \boldsymbol{\varepsilon} \rangle_{V_r} = \langle \mathbf{A} \rangle_{V_r} : \mathbf{E} \quad ; \quad \sum_{r=1}^{n} f_r \langle \mathbf{A} \rangle_{V_r} = \mathbf{I} \tag{14}$$

In the isotropic case $\langle \mathbf{A} \rangle_{V_r}$ is reduced to $\langle \mathbf{A} \rangle_{V_r} = A_r^v \mathbf{J} + A_r^d \mathbf{K}$, where $A_r^v$ and $A_r^d$ are volumetric and deviatoric strain localization coefficients.

For an Eshelbian type morphology [28], i.e. an ellipsoidal inclusion embedded in a reference medium, an estimate of the strain localization tensor of phase *r*, assuming the isotropy of the local and the reference medium is given by [21]:

$$A_r^v = \frac{\left(1 + \alpha_0 \left(k_r/k_0 - 1\right)\right)^{-1}}{\sum_r f_r \left(1 + \alpha_0 \left(k_r/k_0 - 1\right)\right)^{-1}} \tag{15}$$

$$A_r^d = \frac{\left(1 + \beta_0 \left(g_r/g_0 - 1\right)\right)^{-1}}{\sum_r f_r \left(1 + \beta_0 \left(g_r/g_0 - 1\right)\right)^{-1}} \tag{16}$$

with

$$\alpha_0 = \frac{3k_0}{3k_0 + 4g_0} \quad ; \quad \beta_0 = \frac{6(k_0 + 2g_0)}{5(3k_0 + 4g_0)} \tag{17}$$

## 2.2.3. Homogenization

The equations of micro-poroelasticity and the homogenization of the poroelastic properties can be derived on a REV submitted to a homogeneous strain boundary condition and an eigenstress. A detailed derivation of the homogenization equations can be found in Dormieux et al. [24] and Ulm et al. [1]. The main equations are recalled in the following.

The tensor of the overall *effective* moduli of the heterogeneous porous material $\mathbf{C}^{hom}$, is given as:

$$\mathbf{C}^{hom} = \langle \mathbf{c} : \mathbf{A} \rangle_V = \sum_{r=1}^{n} f_r \mathbf{c}_r \langle \mathbf{A} \rangle_{V_r} \tag{18}$$

In the isotropic case $\mathbf{C}^{hom}$ can be presented in the following form:

$$\mathbf{C}^{hom} = 3K_d^{hom} \mathbf{J} + 2G^{hom} \mathbf{K} \tag{19}$$

In the case of an Eshelbian type morphology, the strain concentration coefficients $A_r^v$ and $A_r^d$ can be evaluated using equations (15) and (16). According to the choice of the reference medium in these equations one can distinguish two different homogenization schemes: the *Mori-Tanaka* scheme [29] in which the reference medium is chosen to be the matrix phase; the *Self-consistent* scheme [30] in which the reference medium is the homogenized medium. The Mori-Tanaka scheme is mostly adapted to the composite materials in which the continuous matrix plays a prominent morphological role in the behaviour of the material, e.g. particle reinforced composites for small volume fractions of particles. In this case we have $k_0 = k^{mat}$ and $g_0 = g^{mat}$ where the superscript "mat" stands for the material phase which is considered as the reference medium. The Self consistent scheme is adequate for materials, such as polycrystals, whose phases are dispersed in the RVE so that none of them plays any specific morphological role [21]. The self-consistent scheme is associated with the choice of $k_0 = K_d^{hom}$ and $g_0 = G^{hom}$ which are unknown in advance. Consequently, equation (18) can not be solved directly and the homogenized elastic properties should be calculated using iterative calculations.

The tensor of *effective* Biot coefficients **b** is given by:

$$\mathbf{b}^{hom} = \phi_0 \mathbf{1} : \langle \mathbf{A} \rangle_{V_\phi} = \mathbf{1} : \left( \mathbf{I} - \sum_{r=1}^{m} f_r \langle \mathbf{A} \rangle_{V_r} \right) \tag{20}$$

where $\langle \mathbf{A} \rangle_{V_\phi}$ is the volume average of the strain localization tensor over the pore volume. The tensor of effective solid moduli $\mathbf{C}_s^{hom}$ is given by:

$$\mathbf{C}_s^{hom} = \langle \mathbf{c} : \mathbf{A} \rangle_{V_s} : \langle \mathbf{A} \rangle_{V_s}^{-1} \tag{21}$$

where $\langle \mathbf{A} \rangle_{V_s}$ is the volume average of the strain localization tensor over the solid volume. The *effective* Biot skeleton modulus $N^{hom}$ can be evaluated from the following expression:

$$\frac{1}{N^{hom}} = \mathbf{1} : \sum_{r=1}^{m} f_r \mathbf{c}_r^{-1} : \left( \mathbf{1} - \mathbf{1} : \langle \mathbf{A} \rangle_{V_r} \right) \tag{22}$$

## 2.2.4. Multi-scale porous material

A particular situation, mentioned by Ulm *et al.* [1] and Dormieux *et al.* [24], which is not addressed directly in the standard micro-poroelasticity as presented in the previous section is the case of a porous material in which the pore volume manifests itself at two or several different scales. These two pore volumes are connected and there is one homogeneous pore pressure in all parts of the pore volume. The homogenization of poroelastic properties of such a porous material needs a multi-step homogenization technique. The first step of this procedure is the homogenization of the porous phases which have the smallest pores. This step is performed using the standard homogenization equations as presented in the previous section. The next step of homogenization is concerned with a heterogeneous material composed of some porous phases, some solid phases and a pore volume with a greater length scale than the one inside the porous phases. Let us consider a two-scale porous material in which the pore volume exhibits two different scales I and II ($V_\phi = V_\phi^I + V_\phi^{II}$), i.e. a micro-porosity and a macro-porosity. This material is composed of $l$ ($l \leq m$) porous phases with the porosities $\phi_r^I$, $m-l$ solid phases and a pore volume with the porosity $\phi^{II}$. The total porosity of the material is thus given by:

$$\phi = \sum_{r=1}^{l} f_r \phi_r^I + \phi^{II} \qquad (23)$$

Standard homogenization equations, as presented in the previous section, give the poroelastic properties of the $l$ porous phases of level I ($\mathbf{c}_r^I$, $\mathbf{b}_r^I$, $N_r^I$, $\mathbf{c}_{sr}^I$). The homogenized tensor of Biot effective stress coefficients is found to be:

$$\mathbf{b}^{\text{hom}} = \mathbf{1} - \sum_{r=1}^{m} \left( f_r \langle \mathbf{A} \rangle_{V_r} : \left( \mathbf{1} - \mathbf{b}_r^I \right) \right) \qquad (24)$$

It can be easily verified that equation (24) reduces to equation (20) when all solid phases are non-porous ($\mathbf{b}_r^I = 0$), i.e. the porosity is taking effect in a single length scale. The homogenized Biot modulus is given by the following expression:

$$\frac{1}{N^{\text{hom}}} = \sum_{r=1}^{m} f_r \left( \left( \mathbf{c}_{sr}^I \right)^{-1} : \left( \mathbf{1} - \mathbf{1} : \langle \mathbf{A} \rangle_{V_r} \right) : \left( \mathbf{1} - \mathbf{b}_r^I \right) + \frac{1}{N_r^I} \right) \qquad (25)$$

It can be easily verified that when all solid phases are non-porous ($\mathbf{b}_r^I = 0$, $1/N_r^I = 0$, $\mathbf{c}_{sr}^I = \mathbf{c}_r$), equation (25) reduces to equation (22).

## 3. Experimental evaluation of macroscopic poroelastic parameters

To evaluate the macroscopic poroelastic parameters of the hardened cement paste, an experimental program of drained and undrained isotropic compression tests as well as unjacketed tests has been performed. The results are presented in [4] and [7] and are briefly recalled in the following.

The tests were performed on cylindrical samples with 38mm diameter and 76mm length, made from class G oil well cement at w/c=0.44. The samples have been cured for at least three months in a bath containing an equilibrated fluid under a controlled temperature of 90°C. This temperature was chosen to reproduce the curing conditions of a cement lining installed in a deep (~2 km) oil well.

The porosity of the samples was studied by two methods: oven drying and mercury intrusion porosimetry. The total porosity was measured by drying the samples at 105°C until a constant mass is achieved, and an average value equal to $\phi = 0.35$ is obtained. Mercury intrusion porosimetry was performed with a maximum intruding pressure of 200MPa, and the porosity is equal to $\phi = 0.26$. The maximum intruding pressure of 200MPa corresponds to a minimum pore diameter of about 6nm.

The results of the unjacketed compression test showed that the unjacketed modulus is constant in isothermal conditions. For a porous material with a constant unjacketed modulus, it can be theoretically shown that the stress-dependency of the drained bulk modulus should be controlled by the Terzaghi effective stress [4][12]. To verify this property and to study the simultaneous effects of the pore pressure and of the confining pressure on the behaviour of the material, the drained isotropic compression tests have been performed under different imposed constant pore pressures. Drained and undrained tests were performed with several loading-unloading cycles at different stress levels to study the effect of the effective stress level on the poroelastic properties of the material. The drained tests showed that the stress-dependency of the elastic bulk modulus of the material is indeed controlled by Terzaghi effective stress.

The drained and undrained elastic bulk moduli of the material decrease with Terzaghi effective stress increase. Microscopic observations showed that this degradation is caused by the microcracking of the material even under isotropic loading. This is attributed to the strongly heterogeneous microstructure of the material.

The results of unjacketed and drained tests enable us to evaluate the Biot effective stress coefficient. The analysis of the undrained compression tests in terms of Biot effective stress permits to evaluate the effective drained bulk modulus of the material in undrained tests and to compare them with the results of the drained tests. The comparison of the set of poroelastic parameters obtained from the various independent tests showed a good compatibility and consistency within the framework of the theory of porous media and demonstrated that the behaviour of the hardened cement paste can be indeed described within the framework of this theory.

In this paper the stress-dependency of drained and undrained bulk moduli and the observed microcracking and degradation phenomenon are neglected and the intact values of these parameters, along with the unjacketed modulus and Skempton coefficient are used. The experimental evaluation of these parameters permits the indirect evaluation of the other poroelastic parameters using the relations presented in section 2.1. The complete set of poroelastic parameters is presented in Table (1). As can be seen in this table, a range of values are presented for $K_\phi$, $N$ and $M$. In fact, as mentioned earlier, the modulus $K_\phi$ is very difficult to measure experimentally. On the other hand, the experimental evaluation of the poroelastic parameters $K_d$, $K_s$, $K_u$ and $B$ is more common, so that using these moduli that can be measured independently and equations (8) to (11) one can find four different expressions for indirect evaluation of the parameter $K_\phi$.

$$\frac{1}{K_\phi} = \begin{cases} \dfrac{1}{K_f} - \dfrac{(1/K_d - 1/K_s)(1/K_u - 1/K_s)}{\phi(1/K_d - 1/K_u)} & (a) \\ \dfrac{1}{K_f} - \dfrac{(1-B)(1/K_d - 1/K_s)}{\phi B} & (b) \\ \dfrac{1}{K_f} - \dfrac{1/K_u - 1/K_s}{\phi B} & (c) \\ \dfrac{1}{K_f} - \dfrac{(1-B)(1/K_d - 1/K_u)}{\phi B^2} & (d) \end{cases} \quad (26)$$

The numerical value of the modulus $K_\phi$ is very sensitive to small changes of each of the other parameters in equation (26). Consequently, despite of the very good compatibility of the experimentally evaluated parameters as demonstrated in [4], the indirect evaluation of the modulus $K_\phi$ using different expressions in equation (26) results in values which vary between 13.0 and 21.9 GPa. This variability of the evaluated $K_\phi$ induces a variability in the evaluation of $N$ and $M$ using equations (8) and (9) respectively.

| Experimentally evaluated parameters | | | | |
|---|---|---|---|---|
| $K_d$ (GPa) | $K_s$ (GPa) | $K_u$ (GPa) | $B$ (-) | $G$ (GPa) |
| 8.7 | 21.0 | 11.2 | 0.4 | 5.7 |
| **Indirectly evaluated parameters** | | | | |
| $b$ (-) | $K_p$ (GPa) | $K_\phi$ (GPa) | $N$ (GPa) | $M$ (GPa) |
| 0.59 | 3.85 | 13.0 - 21.9 Ave=16.9 | 62 - 126 Ave=80 | 7.5 - 7.9 Ave=7.7 |

**Table (1): Experimentally evaluated poroelastic parameters of the tested hardened cement paste**

## 4. Microstructure of cement paste

The cement clinker is composed of four main phases: $C_3S$, $C_2S$, $C_3A$ and $C_4AF$ where in the standard cement chemistry the notation C stands for CaO, S for $SiO_2$, A for $Al_2O_3$ and F for $Fe_2O_3$. The setting and hardening of the cement paste are the results of the complex reactions, called hydration reactions, between the clinker phases and the water. In a classification initiated by Taplin [31], the hydration products can be designated as either outer (Op) or inner (Ip) products; outer products form in the originally water-filled spaces and inner products within the boundaries of the original clinker grains. But there is not necessarily an exact correspondence between the positions of the outer boundaries of Ip and original grains [32]. The cement paste has a very complex microstructure which varies with the cement composition, time and hydration conditions. In a simplified view, the main phases of the microstructure are calcium-silicate-hydrate (C–S–H) which is the main binding phase of all Portland cement-based materials, Portlandite (CH), Aluminates (AL), cement clinkers and macro-porosity. C−S−H is the main hydration product which is a porous phase with an amorphous and colloidal structure and a variable chemical composition. The CH often occurs as massive crystals but is also mixed with C−S−H at the micron-scale. CH and cement clinker can be considered as non-porous solid phases.

Because of its colloidal and amorphous nature and the variability of its chemical composition, the structure of C−S−H matrix and its solid phase are not clearly known. Since a few decades, starting from the work of Powers and Brownyard [33] who recognized the colloidal properties of C−S−H, different models have been proposed in the literature for the structure of this material. Most of these models, like the one proposed by Feldman and Sereda [34], consider a layered structure for C−S−H and also the existence of an important quantity of chemically bonded or adsorbed water. Jennings [35][36] proposed a microstructural model for C−S−H in which the amorphous and colloidal structure of the C−S−H is organized in elements, called 'globules'. The globule, with a size of about 4nm, is composed of solid C−S−H sheets, intra-globule porosity filled with structural water and a monolayer of water on the surface. The structure of C−S−H in Jennings' model contains small gel pores in the space between adjacent globules and larger gel pores between the groups of several globules. Jennings' model distinguishes two types of C−S−H, called low density and high density C−S−H. The globules are considered to be identical in LD and HD C−S−H and the difference between these two types of C−S−H is in the gel porosity of about 0.24 for HD C−S−H and 0.37 for LD C−S−H. The classification of HD and LD C−S−H in Jennings' model is more or less equivalent to the Ip and Op C−S−H classification. Richardson [32][37] studied the composition, morphology, and spatial distribution of various hydration products, mainly based on the high resolution images taken by transmission electron microscopy (TEM) on very thin sections, and showed that a clear distinction can be made between the morphology of Ip and Op C−S−H. According to Richardson, Op C−S−H has a fibrillar, directional morphology which is a function of space constraint: where it forms in large pore spaces, the fibrils form with a high aspect ratio ('coarse fibrillar'), while in smaller spaces, it retains a directional aspect but forms in a more space-efficient manner ('fine fibrillar'). Ip C−S−H has a compact, fine-scale and homogeneous morphology, with only gel porosity, i.e. pores within the C−S−H, which are approximately smaller than 10nm. Constantinides and Ulm [38][39] studied the mechanical behaviour of C−S−H using the nanoindentation tests. The statistical analysis of hundreds of nanoindentation tests showed the existence of two distinguishable peaks of indentation modulus which verifies the existence of two structurally distinct but compositionally similar types of C−S−H. These authors show that the C−S−H has a nanogranular behaviour which is driven by particle-to-particle contact forces. Constantinides and Ulm [39] attributed the packing densities of HD and LD C−S−H to two limit values which correspond to the random packing limit in the case of LD C−S−H.

For the purpose of micromechanics modelling and homogenization of poroelastic properties, the microstructure of the hardened cement paste is divided into the following three scale levels:

- Level 0 ($10^{-9}$–$10^{-8}$m, the C−S−H solid): Solid phase of C−S−H matrix.

- Level 1 ($10^{-8}$–$10^{-6}$m, the C–S–H matrix): High density and low density C–S–H.
- Level 2 ($10^{-6}$–$10^{-4}$m, the cement paste): C–S–H matrix, CH, AL, cement clinker and water.

This multi-scale microstructure is used for homogenization of the poroelastic parameters of the cement paste presented in section 5.

## *4.1. Cement paste active porosity in poromechanics*

The analysis of the experimentally evaluated poroelastic parameters of the hardened cement paste in Ghabezloo *et al*. [4] revealed that the active porosity of the tested cement paste in the performed tests is smaller than its total porosity. This statement is based on the following inequality from the theory of poroelasticity [12]:

$$\frac{b}{K_s} - \frac{\phi}{K_\phi} \geq 0 \qquad (27)$$

Note that in the case of a micro-homogeneous porous material for which $K_s = K_\phi$, equation (27) is reduced to the well-known inequality $b \geq \phi$ in poroelasticity theory. Using equations (6) and (27) in equations (8) to (11) the following upper limit is obtained for the porosity as a function of other poroelastic parameters:

$$\phi \leq K_f \left( \frac{b}{K_s} + \frac{1/K_u - 1/K_s}{B} \right) \qquad (28)$$

Replacing the values of the parameters presented in Table (1), $K_u = 11.25\text{GPa}$, $K_s = 21\text{GPa}$, $b = 0.586$, $B = 0.4$, and taking $K_f = 2.2\text{GPa}$, from equation (28) we obtain $\phi \leq 0.29$ which is smaller than the total porosity of the tested cement paste, evaluated equal to 0.35. The difference between the total porosity and the porosity upper limit means that a part of the porosity of was not active in the laboratory experiments. A pore volume can be considered to be active if under the effect of a pressure gradient, the pore fluid can exchange with the fluid filling the pore volume situated in its neighbourhood. There are situations in which the pore fluid mass exchange is limited, as for example due to the:

- lack of connection between the considered pore volume and the pore volume in its neighbourhood (occluded porosity).
- adsorbing forces applied to the fluid in the close vicinity of the solid particles. These forces reduce the mobility of water molecules [40].
- very low permeability of the considered pore volume which does not permit fluid mass exchange to take place in the limits of the time-scale of the applied load for the length of the drainage path.

The difference between the total porosity and the upper limit for active porosity can thus be attributed partly to each of the above mentioned factors. From the poromechanics point of view, the inactive pore volume and the pore fluid filling it should be considered as a part of the solid phase.

From different microstructural models of C–S–H, it can be seen that a part of the water in the pore structure of cement paste is interlayer structural water. The model proposed by Feldman and Sereda [33] for multilayer structure of C−S−H, postulates the existence of interlayer space containing strongly adsorbed water. According to Feldman [41], the interlayer water behaves as a solid bridge between the layers and consequently the interlayer space can not be included in the porosity and the interlayer water must be regarded as a part of the solid structure of hydrated cement paste. This is compatible with the results of molecular dynamic simulations of Pellenq et al. [42] which show that the diffusion of water molecules is about a thousand times slower in the interlayer space than in bulk water at the same temperature. Feldman's experiments show that the interlayer water evaporates at very low relative humidity, below 11% [41]. This can be seen also in Jennings' [35][36] microstructural model of C−S−H. In this model for relative humidity below 11%, a part of the water filling the intra-globule porosity is evaporated. Accordingly, in a porosity measurement in which the sample is dried at 105°C until a constant mass is achieved, a part of the interlayer water is evaporated and is thus included in the measured porosity [43][44]. The free-water porosity is defined and measured by equilibrating the cement paste at 11% relative humidity [43][44] and is obviously smaller than the total porosity which is measured by drying at 105°C. Some measurements of the free-water porosity from Feldman [45] are presented by Jennings et al. [44]. Based on the above discussion, Ghabezloo et al. [4] argued that the free-water porosity is the cement paste porosity that should be used in the poromechanical calculations. Due to the lack of direct measurements of the free water porosity, Ghabezloo et al. [4] approximated the free-water porosity by the porosity obtained by mercury intrusion. This approximation was based on some data provided by Taylor [43] (in Fig. 8.5), on the experimentally evaluated values of free-water porosity and the mercury porosity. Scherer et al. [40] also argued that a part of the pore fluid in the microstructure of the cement paste is inactive. According to these authors, on the surface of the solid particles, the thickness of the layer of immobile water is about 0.5nm which corresponds to two water molecules. Accordingly, these authors reduce the total porosity of the cement paste for the effect of the thickness of the immobile water (see equation 17 in [40]). In a similar approach, Sun and Scherer [46][47] excluded the volume of interlayer water in evaluation of the active porosity of mortar samples.

The distribution of the active pore volume within the total pore volume of the cement paste is not accurately known, but is important for the homogenization of the poroelastic properties. Among the three above mentioned causes (occluded pores, surface forces on water molecules, low permeability zones) of the inactive porosity, the distribution of the effect of the adsorbing forces of solid particles in the microstructure naturally follows the distribution of the solid particles. It is known that the C−S−H solid has a greater packing density in HD C−S−H than in LD C−S−H, consequently the inactive porosity resulting from the adsorbing forces of solid particles on the water molecules in their close vicinity is mostly concentrated in the HD C−S−H. For the same reason, the permeability of HD C−S−H should be considerably lower than the one of the LD C−S−H. Consequently, the inactive porosity resulting from the very low permeability zones is also situated mostly in HD C−S−H. Accordingly, a reasonable assumption is that the inactive porosity is entirely situated in HD C−S−H. Consequently, the active porosity of the cement paste consists of the porosity in LD C−S−H and the macro-porosity. This assumption is similar to the one made by Tennis and Jennings [48] for evaluation of the volume fraction of LD C−S−H in the microstructure of the hardened cement paste by analysing the results of surface area measurements by nitrogen sorption. These authors assumed that none of the pores in HD C−S−H are accessible to nitrogen. The empirical relation obtained using this assumption for the volume fraction of LD C−S−H is widely used in the literature.

Assuming that the porosity in HD C−S−H is not active from the poromechanics point of view means that in the time-scale of the applied loads, the mass of the pore fluid in HD C−S−H porosity is constant. Consequently the HD C−S−H behaves like a porous material in undrained conditions. This consideration has some consequences for the equations of the homogenization model for the poroelastic properties of the hardened cement paste and is discussed in section 5.1.

## *4.2. Volume fractions*

The *description* step of the homogenization method needs the evaluation of the volume fractions of different constituents of the cement paste. For an ordinary Portland cement paste hydrated at ambient temperature, these volume fractions can be evaluated using the Power's hydration model [33] as a function of the water-to-cement ratio and the hydration degree. Because of the simplicity of its implementation, this model is widely used in the homogenisation models for the properties of the cement paste, as for example by Sanahuja *et al.* [2] and Pichler *et al.* [3]. A more advanced hydration model is presented by Jennings and Tennis [49] which takes into account the chemical composition of the cement clinker for the evaluation of the volume fractions. As mentioned before, the cement paste used in this study is prepared using class G cement. For this cement paste, the volume fractions of the different phases of the microstructure can be evaluated as functions of cement composition, water-to-cement ratio and degree of hydration, using the method presented by Bernard *et al.* [50]. This model assumes simple stoichiometric reactions for the hydration of the four dominant compounds in Portland cement, $C_3S$, $C_2S$, $C_3A$ and $C_4AF$. The complete set of chemical reactions is presented in [48]. The following relations show the production of C−S−H and CH during the hydration of $C_3S$ and $C_2S$:

$$2C_3S+10.6H \rightarrow C_{3.4}\text{-}S_2\text{-}H_8+2.6CH$$
$$2C_2S+8.6H \rightarrow C_{3.4}\text{-}S_2\text{-}H_8+0.6CH$$
(29)

In addition to the four principal phases mentioned above with mass fractions $m_i$, normally the cement contains some quantity of gypsum and impurities and so the sum of mass fractions $m_i$ for the four principal phases is smaller than one. The normalized mass fraction $m'_i$ is defined as:

$$m'_i = \frac{m_i}{\sum_i m_i} \qquad (i = C_3S, C_2S, C_3A, C_4AF)$$
(30)

The total volume of the cement paste at a given time is composed of the volume of the reactants (remaining water and cement grains) and products (C−S−H, Aluminates, CH, capillary voids):

$$V_{cp}^{total} = \underbrace{V_w + \sum_i V_i^{ck}}_{\text{Reactants}} + \underbrace{V_{CSH} + V_{CH} + V_{AL} + V_{Cap}}_{\text{Products}}$$
(31)

where $V_w$ is the volume of remaining water and $V_i^{ck}$ ($i$=C$_3$S, C$_2$S, C$_3$A, C$_4$AF) is the volume of remaining clinker phase $i$. The hydration degree $\alpha_i$ is defined as the ratio of the remaining mass of the clinker phase $i$ to its initial mass. The overall hydration degree is defined as [43]:

$$\alpha = \sum_i m_i' \alpha_i \qquad (32)$$

The volume of the remaining water is obtained by subtraction of the water consumed during hydration of clinker phases from the initial water content $V_w^0$.

$$V_w = V_w^0 - \sum_i V_w^i \alpha_i \qquad (33)$$

$\alpha_i$ is the degree of hydration of clinker phase $i$ and $V_w^i$ is the volume of consumed water for complete hydration of this phase.

$$\frac{V_w^i}{V_c^0} = N_w^i \frac{\beta_i^*}{\beta_w} \quad ; \quad \beta = \frac{\rho}{\mu} \quad ; \quad \rho_i^* = \rho_c m_i' \qquad (34)$$

$V_c^0$ is the initial cement volume. The parameter $\beta$ represents the number of moles (of clinker phase $i$, water, CH or CSH) per unit volume of cement, and is defined as the ratio of the molar mass $\mu$ to the mass density $\rho$. $\rho_c$ is the cement density and $N_w^i$ is the number of moles of the consumed water during the hydration of one mol of clinker phase $i$. The volume of produced C–S–H is given by:

$$V_{CSH} = \sum_i V_{CSH}^i \alpha_i \quad ; \quad \frac{V_{CSH}^i}{V_c^0} = N_{CSH}^i \frac{\beta_i^*}{\beta_{CSH}} \quad (i = C_2S, C_3S) \qquad (35)$$

where $V_{CSH}^i$ is the volume of C–S–H produced by complete hydration of clinker phase $i$. Similarly, the volume of produced CH is given by the following relation:

$$V_{CH} = \sum_i V_{CH}^i \alpha_i \quad ; \quad \frac{V_{CH}^i}{V_c^0} = N_{CH}^i \frac{\beta_i^*}{\beta_{CH}} \quad (i = C_2S, C_3S) \qquad (36)$$

The volumes of remaining clinker phases are calculated using the relation below:

$$\frac{V_{CK}^i}{V_c^0} = m_i'(1 - \alpha_i) \qquad (37)$$

The capillary voids are produced by the chemical shrinkage of the hydrates during the hydration:

$$\frac{V_{cap}}{V_c^0} = c_{sr} \rho_c \alpha \qquad (38)$$

where $c_{sr}$ is the chemical shrinkage per gram of cement at complete hydration for which the value of 0.07 cm$^3$/g is used [51][52]. The volume fractions of phases of the microstructure of the cement paste can be calculated from the below relation in which it is assumed that the hydration is performed with constant total volume:

$$f_i = \frac{V_i}{V_{cp}^{total}} = \frac{V_i/V_c^0}{1 + V_w^0/V_c^0} = \frac{V_i/V_c^0}{1 + \frac{\rho_c}{\rho_w} \frac{w}{c}} \qquad (39)$$

The volume fraction of aluminates is calculated from:

$$f_{AL} = 1 - \left( f_{CSH} + f_{CH} + f_w + f_{cap} + \sum_i f_i^{ck} \right) \tag{40}$$

The remaining water and capillary voids from the macro-porosity of the cement paste:

$$f_V = f_w + f_{cap} \tag{41}$$

The values of model parameters along with the composition of the class G cement used in this study are presented in Table (2). The model evaluates the total volume fraction of C−S−H phase, but does directly provide the volume fractions of HD and LD C−S−H. These volume fractions can be estimated by assuming, as proposed by Bernard *et al*. [50], that the LD C−S−H corresponds to the outer products and the HD C−S−H to the inner products. Moreover, the outer products are formed during the nucleation and growth phase of the hydration process ($\alpha_i \leq \alpha_i^*$) and the inner products during the diffusion-controlled phase ($\alpha_i > \alpha_i^*$). During the hydration process, beyond a critical hydration degree $\alpha_i^*$ which corresponds to a critical thickness of hydration products formed around the clinker grains, the kinetics of the hydration reactions is limited by diffusion of dissolved ions through the layers of hydrates formed around the clinker. Bernard *et al*. [50] take $\alpha_i^* = 0.6$ for all clinker phases and the same value is taken here.

| Parameters | Reactants | | | | | | Products | |
|---|---|---|---|---|---|---|---|---|
| | $C_3S$ | $C_2S$ | $C_3A$ | $C_4AF$ | w | c | $C_{3.4}$–$S_2$–$H_8$ | CH |
| $\rho_i$ (g/cm³) | - | - | - | - | 1.00 | 3.15 | 2.04 | 2.24 |
| $m_i$ | 0.63 | 0.14 | 0.02 | 0.13 | - | - | - | - |
| $\mu_i$ (g/mol) | 228.32 | 172.24 | 270.20 | 485.96 | 18 | - | 454.95 | 74 |
| $N_{CSH}^i$ | 0.5 | 0.5 | - | - | - | - | - | - |
| $N_{CH}^i$ | 1.3 | 0.3 | - | - | - | - | - | - |
| $N_w^i$ | 5.3 | 4.3 | 10.0 | 10.75 | - | - | - | - |

**Table (2): Hydration model parameters and the composition of the class G cement used in the experimental study**

The volumes of LD and HD C−S−H are given by the following relations:

$$V_{CSH}^{LD} = \sum_i V_{CSH}^i \left( \alpha_i^* - \langle \alpha_i^* - \alpha_i \rangle_+ \right) \quad (i = C_3S, C_2S) \tag{42}$$

$$V_{CSH}^{HD} = \sum_i V_{CSH}^i \left( \langle \alpha_i - \alpha_i^* \rangle_+ \right) \quad (i = C_3S, C_2S) \tag{43}$$

where $\langle z \rangle_+$ is defined as:

$$\langle z \rangle_+ = \begin{cases} z & \text{if } z > 0 \\ 0 & \text{if } z \leq 0 \end{cases} \tag{44}$$

The volume fraction of the LD C−S−H in the C−S−H matrix is given by:

$$f_{CSH}^{LD} = \frac{V_{CSH}^{LD}}{V_{CSH}^{HD} + V_{CSH}^{LD}} \tag{45}$$

The volume fractions of LD and HD C−S−H in the cement paste can be calculated from the following relations:

$$f_{LD} = f_{CSH}^{LD} f_{CSH} \quad ; \quad f_{HD} = \left(1 - f_{CSH}^{LD}\right) f_{CSH} \tag{46}$$

Though the method proposed above provides an estimate of the volume fractions of the C–S–H phases, it cannot completely reflect the dependence of these volume fractions on the hydration conditions. These limitations are partly due to the assumptions, as mentioned above, and partly due to the lack of information about the dependence of the critical hydration degree $\alpha_i^*$ on the hydration conditions and clinker properties. By assuming a constant value for $\alpha_i^*$, the volume fraction of LD C–S–H for complete hydration of any cement paste and any hydration temperature using the presented method is found equal to 0.6. This is not compatible with some observations, as for example the results of nanoindentation tests of Constantinides [53] which show the reduction of the volume fraction of LD C–S–H by increasing the hydration temperature. However, the knowledge of the dependence of the volume fractions of different phases of C–S–H matrix on hydration conditions is still limited and consequently in this work we adopt the method presented here above for evaluation of the volume fractions of these phases.

Since the hardened cement paste used in the experimental program was cured for at least 3 months at 90°C before performing the tests, we can assume that it is completely hydrated. Table (3) presents the volume fractions of the hardened cement paste that are calculated by assuming $\alpha_i = 1.0$ for all clinker phases. These values will be used for the calibration of the homogenization model, as presented in section 5.2. Considering the age of the studied cement paste, the assumption of complete hydration has a negligible effect on the modelling results. Alternatively, one can evaluate the hydration kinetics by using an existing model, as for instance the one presented by Bernard *et al.* [50], and calculate the hydration degree $\alpha_i$ of each clinker phase. Evaluation of the hydration kinetics will be necessary for evaluation of the poroelastic properties during the hydration process using the micromechanics model.

| Constituents | C–S–H | CH | CK | AL | w | cap |
|---|---|---|---|---|---|---|
| $f_i$ | 0.58 | 0.18 | 0.00 | 0.13 | 0.02 | 0.09 |

**Table (3): Evaluated volume fractions for the microstructure of the tested hardened cement paste.**

## *4.3. Material properties*

The development of advanced micromechanical testing methods, particularly the nano-indentation tests, has provided direct measurements of the elastic properties of the different phases of the microstructure of the hardened cement paste. The mechanical properties of cement clinker phases have been the object of several studies. When measured by nanoindentation and resonance frequency, Young's moduli of $C_3S$, $C_2S$, $C_3A$ and $C_4AF$ are approximately similar, with values between 125 and 150 GPa [54][55]. The Poisson's ratios for these phases are similar, equal to 0.3 [54][55]. Atomistic calculations recently presented by Manzano *et al.* [56] resulted in Young's moduli of $C_3S$ and $C_2S$ equal to about 140 GPa. The Poisson's ratios for these clinker phases are 0.28 and 0.3 respectively. Nanoindentation measurements give values between 36 and 40

GPa for Young's modulus of Portlandite [38][54]. These results are in agreement with the values obtained by extrapolation to zero porosity, equal to 35 GPa and 48 GPa, as presented respectively in [57] and [58]. Monteiro and Chang [59] evaluated the Poisson's ratio of Portlandite by extrapolation to zero porosity, equal to 0.305. By means of atomistic calculations, Manzano *et al*. [56] evaluated Young's modulus and Poisson's ratio of Portlandite respectively equal to 35.2 GPa and 0.31, which are compatible with the above mentioned values. Unfortunately, there is a lack of data on the elastic properties of aluminates, so we assume that they have the same mechanical properties as C–S–H phase.

# 5. Homogenization of poroelastic properties of the hardened cement paste

The classical homogenization procedure, used for example by Ulm *et al*. [1], is to evaluate the microscopic properties, for example by nanoindentation testing, and then calculate the homogenized macroscopic poroelastic properties of the hardened cement paste using these parameters. In the present work, the macroscopic properties are already known for a cement paste with a fixed water-to-cement ratio. The calibration of the unknown microscopic parameters from these results will then permit us to evaluate the same macroscopic poroelastic properties for a cement paste having different water-to-cement ratio.

## *5.1. Homogenization equations*

Considering the multi-scale microstructure of the cement paste presented in section 4, the homogenization of the macroscopic poroelastic properties is done here in two steps. The first step deals with the homogenization of the poroelastic properties of high density and low density C–S–H. The macroscopic poroelastic properties of the cement paste are then evaluated in the next homogenization step using the properties of C–S–H evaluated in the first step and also the properties of the other phases of the microstructure. Figure (1) gives a micromechanics representation of the REVs of C–S–H and cement paste which are used in the homogenization model. High density and low density C–S–H are composed of C–S–H solid particles and gel pores, with a higher porosity in the case of LD C–S–H. The cement paste microstructure is composed of LD C–S–H, HD C–S–H, cement clinker and macroporosity. Considering the self-consistent homogenization scheme, the matrix in the C–S–H phase and in the cement paste is equivalent to the homogenized medium.

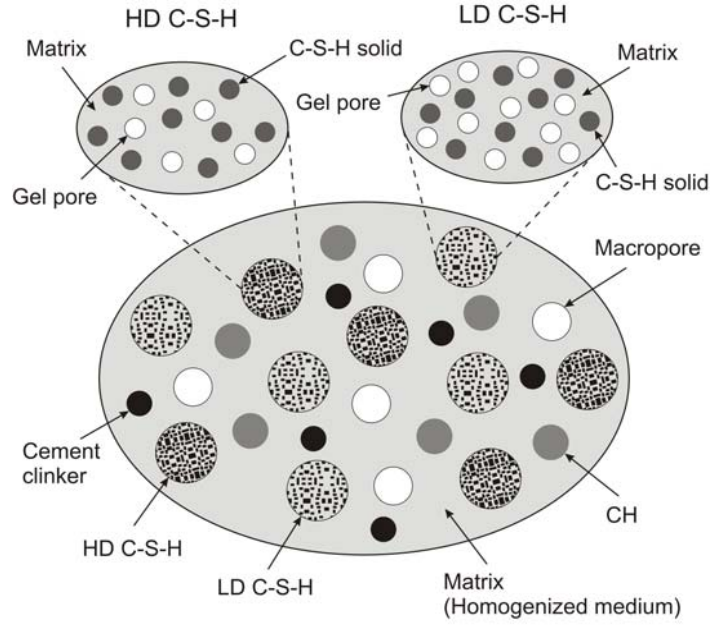

**Figure (1): Micromechanics representation of the cement paste REV**

### 5.1.1. Level 1: C−S−H matrix

The only difference between HD and LD C−S−H is in their packing density or porosity. The homogenization of poroelastic properties of C−S−H matrix can be done using equations presented in section 2.2.3. The needed parameters are the porosities, $\phi_{LD}$ and $\phi_{HD}$, and the elastic parameters of the C−S−H solid, $k_s$ and $g_s$. The expressions of the homogenized poroelastic parameters of high density and low density C−S−H are presented in the following equations, in which the subscript X represents LD or HD.

$$K_X^{hom} = (1-\phi_X)k_s A_{s,X}^v \tag{47}$$

$$G_X^{hom} = (1-\phi_X)g_s A_{s,X}^d \tag{48}$$

$$b_X^{hom} = 1 - \frac{K_X^{hom}}{k_s} \tag{49}$$

$$\frac{1}{N_X^{hom}} = \frac{(1-\phi_X)(1-A_{s,X}^v)}{k_s} \tag{50}$$

Assuming an Eshelbian type morphology, the strain localization tensor parameters, $A_{r,X}^v$ and $A_{r,X}^d$, of each phase can be estimated using equations (15) and (16) considering a solid and a porous phase, $(r = s, \phi)$. As argued by Stora et al. [60], considering spherical inclusions for the homogenization of elastic parameters of sound hardened cement paste is a suitable assumption. Considering the self-consistent scheme we should take $k_0 = K_X^{hom}$, $g_0 = G_X^{hom}$ and the parameters $\alpha_0$ and $\beta_0$ are given by the following relations:

$$\alpha_0 = \frac{3K_X^{hom}}{3K_X^{hom}+4G_X^{hom}} \quad;\quad \beta_0 = \frac{6(K_X^{hom}+2G_X^{hom})}{5(3K_X^{hom}+4G_X^{hom})} \tag{51}$$

As mentioned in section 4.1, we assume that the porosity in HD C–S–H is not active so that the HD C–S–H behaves like a porous material in the undrained condition. Consequently, in the second homogenization step the undrained bulk modulus of HD C–S–H should be used. From equations (9), (10) and (11) the following relation can be obtained for evaluation of the undrained bulk modulus of the HD C–S–H. The pore fluid bulk modulus $K_f$ can be taken equal to 2.2 GPa at ambient temperature.

$$K_{u,\text{HD}}^{\text{hom}} = K_{\text{HD}}^{\text{hom}} + \frac{\left(b_{\text{HD}}^{\text{hom}}\right)^2}{\dfrac{1}{N_{\text{HD}}^{\text{hom}}} + \dfrac{\phi_{\text{HD}}}{K_f}} \tag{52}$$

## 5.1.2. Cement paste

The microstructure of the cement paste for the second homogenization step is constituted by five main phases: high density and low density C–S–H, Portlandite, anhydrous clinker and the macro-porosity. These phases are respectively represented by HD, LD, CH, CK and V subscripts. Among these phases, Portlandite crystals and anhydrous clinker can be considered as non-porous solids, while low density and high density C–S–H are porous solids. However we assumed that in the poromechanics tests the porosity in HD C–S–H was not active and so the only porous phase in the microstructure is the LD C–S–H. Consequently the homogenization of the poroelastic properties of the cement paste at this step should be performed using the framework presented in section 2.2.4 for multiphase porous materials, as presented in the following equations:

$$K_{\text{CP}}^{\text{hom}} = f_{\text{LD}} K_{\text{LD}}^{\text{hom}} A_{\text{LD,CP}}^{v} + f_{\text{HD}} K_{u,\text{HD}}^{\text{hom}} A_{\text{HD,CP}}^{v} + f_{\text{CH}} k_{\text{CH}} A_{\text{CH,CP}}^{v} + f_{\text{CK}} k_{\text{CK}} A_{\text{CK,CP}}^{v} \tag{53}$$

$$G_{\text{CP}}^{\text{hom}} = f_{\text{LD}} G_{\text{LD}}^{\text{hom}} A_{\text{LD,CP}}^{d} + f_{\text{HD}} G_{\text{HD}}^{\text{hom}} A_{\text{HD,CP}}^{d} + f_{\text{CH}} g_{\text{CH}} A_{\text{CH,CP}}^{d} + f_{\text{CK}} g_{\text{CK}} A_{\text{CK,CP}}^{d} \tag{54}$$

$$b^{\text{hom}} = 1 - f_{\text{LD}} A_{\text{LD,CP}}^{v} \left(1 - b_{\text{LD}}^{\text{hom}}\right) - f_{\text{HD}} A_{\text{HD,CP}}^{v} - f_{\text{CH}} A_{\text{CH,CP}}^{v} - f_{\text{CK}} A_{\text{CK,CP}}^{v} \tag{55}$$

$$\frac{1}{N^{\text{hom}}} = f_{\text{LD}} \left( \frac{\left(1 - A_{\text{LD,CP}}^{v}\right)\left(1 - b_{\text{LD}}^{\text{hom}}\right)}{k_s} + \frac{1}{N_{\text{LD}}^{\text{hom}}} \right) + f_{\text{HD}} \frac{\left(1 - A_{\text{HD,CP}}^{v}\right)}{K_{u,\text{HD}}^{\text{hom}}}$$
$$+ f_{\text{CH}} \frac{\left(1 - A_{\text{CH,CP}}^{v}\right)}{k_{\text{CH}}} + f_{\text{CK}} \frac{\left(1 - A_{\text{CK,CP}}^{v}\right)}{k_{\text{CK}}} \tag{56}$$

The strain localization tensor parameters, $A_{r,\text{CP}}^{v}$ and $A_{r,\text{CP}}^{d}$, of each phase can be estimated using equations (15) and (16) considering four solid phases and a pore volume, $(r = \text{HD}, \text{LD}, \text{CH}, \text{CK}, \phi)$. For the HD C–S–H the homogenized undrained bulk modulus $k_{u,\text{HD}}^{\text{hom}}$ should be used in equation (15). Considering the self-consistent scheme we should take $k_0 = K_{\text{CP}}^{\text{hom}}$, $g_0 = G_{\text{CP}}^{\text{hom}}$ and the parameters $\alpha_0$ and $\beta_0$ are given by the following relations:

$$\alpha_0 = \frac{3K_{\text{CP}}^{\text{hom}}}{3K_{\text{CP}}^{\text{hom}} + 4G_{\text{CP}}^{\text{hom}}} \quad ; \quad \beta_0 = \frac{6\left(K_{\text{CP}}^{\text{hom}} + 2G_{\text{CP}}^{\text{hom}}\right)}{5\left(3K_{\text{CP}}^{\text{hom}} + 4G_{\text{CP}}^{\text{hom}}\right)} \tag{57}$$

The total porosity of the cement paste is the sum of the porosity of C−S−H matrix and the macro-porosity, as presented in the following expression:

$$\phi_{CP} = f_{LD}\phi_{LD} + f_{HD}\phi_{HD} + f_V \tag{58}$$

As we assumed that the porosity in HD C−S−H is not active from the poromechanics point of view, the active porosity of the cement paste $\phi_{CP}^{act}$ should be calculated by the following expression:

$$\phi_{CP}^{act} = f_{LD}\phi_{LD} + f_V \tag{59}$$

By evaluation of the homogenized macroscopic poroelastic parameters presented in equations (53) to (56) and the active porosity, the remaining poroelastic parameters of the cement paste ($K_u$, $K_s$, $K_\phi$, $B$, $M$) can be evaluated using relations presented in section 2.1.

## 5.2. Model calibration

Table (4) summarizes the parameters for the homogenization of the poroelastic properties of the cement paste. We assume that the Aluminates phase has the same properties as the C−S−H phase, consequently the volume fraction of C−S−H in Table (4) is equal to the sum of volume fractions of Aluminates and C−S−H in Table (3). The properties of the phases are the same as the ones used by Ulm *et al*. [1]. The volume fractions and porosities presented in Table (4) can be used in equation (59) to calculate the active porosity of the hardened cement paste. This gives a value equal to 0.27 which is very close to the porosity evaluated by mercury intrusion, equal to 0.26. This is compatible with the approximation made in Ghabezloo et al. [4][6] by taking the active porosity equal to the mercury porosity.

| C−S−H level | | Cement paste level | |
|---|---|---|---|
| **Parameter** | **Value** | **Parameter** | **Value** |
| $k_s$ | calibrated | $f_{CSH}$ | 0.71 |
| $g_s$ | calibrated | $f_{CH}$ | 0.18 |
| $\phi_{HD}$ | 0.24 | $f_V$ | 0.11 |
| $\phi_{LD}$ | 0.37 | $f_{CK}$ | 0.00 |
| | | $f_{CSH}^{LD}$ | 0.60 |
| | | $k_{CH}$ | 32.5 GPa |
| | | $g_{CH}$ | 14.6 GPa |
| | | $k_{CK}$ | $C_3S$ : 112.5 GPa<br>$C_2S$ : 116.7 GPa<br>$C_3A$ : 120.8 GPa<br>$C_4AF$ : 104.2 GPa |
| | | $g_{CK}$ | $C_3S$ : 51.9 GPa<br>$C_2S$ : 53.8 GPa<br>$C_3A$ : 55.8 GPa<br>$C_4AF$ : 48.1 GPa |

**Table (4): Homogenization model parameters**

As can be seen in Table (4), the two principal unknown parameters are the elastic properties of the C−S−H solid which should be calibrated based on the experimentally evaluated macroscopic parameters presented in Table (1). This is done by minimizing the error between the experimentally evaluated parameters and the results of the homogenization model using a least-squares method. The error is defined by the following relation:

$$E_r = \left(\frac{K_d^{exp} - K_d^{hom}}{K_d^{exp}}\right)^2 + \left(\frac{K_s^{exp} - K_s^{hom}}{K_s^{exp}}\right)^2 + \left(\frac{K_u^{exp} - K_d^{hom}}{K_u^{exp}}\right)^2 + \left(\frac{B^{exp} - B^{hom}}{B^{exp}}\right)^2 + \left(\frac{G^{exp} - G^{hom}}{G^{exp}}\right)^2 \quad (60)$$

The calibration of the elastic parameters of the C–S–H solid is done using a computer program which calculates the homogenized poroelastic properties of the hardened cement paste for different combinations of $k_s$ and $g_s$ and evaluates the error using equation (60). The bulk modulus $k_s$ varied between 20 and 30 GPa, and the shear modulus $g_s$ of the solid phase varied between 13 and 23 GPa. The contour plot of the calculated error for different values of $k_s$ and $g_s$ is presented in Figure (2). The minimum error is found equal to $6.1 \times 10^{-4}$ for the elastic parameters of the C–S–H solid presented in Table (5). The calculated poroelastic parameters of the hardened cement paste are presented in Table (6) and compared with the experimentally evaluated parameters, where we can see a very good compatibility. The calculated Young's modulus of the C–S–H solid, equal to 44.3 GPa, is close to the value obtained by Ulm *et al.* [1], equal to 47.7 GPa, from back analysis of the results of the nanoindentation tests using the Mori-Tanaka homogenization scheme. The calculated Young's modulus is also compatible with the results of the atomistic calculations of Manzano *et al.* [56], which are between 35 and 56 GPa.

| Parameter | $k_s$ (GPa) | $g_s$ (GPa) | $e_s$ (GPa) | $\nu_s$ (-) |
|---|---|---|---|---|
| **Value** | 25.0 | 18.4 | 44.3 | 0.204 |

Table (5): Calibrated elastic parameters of the C–S–H solid

| Parameter | $K_d$ (GPa) | $K_s$ (GPa) | $K_u$ (GPa) | $B$ (-) | $G$ (GPa) |
|---|---|---|---|---|---|
| **Experimental** | 8.7 | 21.0 | 11.2 | 0.40 | 5.7 |
| **Model calibration** | 8.5 | 21.3 | 11.2 | 0.39 | 5.8 |

Table (6): Comparison of the calculated and experimentally evaluated poroelastic parameters of the tested hardened cement paste

Now we have all necessary ingredients to use the homogenization model for extrapolation of the experimentally evaluated poroelastic parameters of the hardened cement paste. This extrapolation is done for the poroelastic properties of hardened cement pastes with different water-to-cement ratios and is presented in the following section.

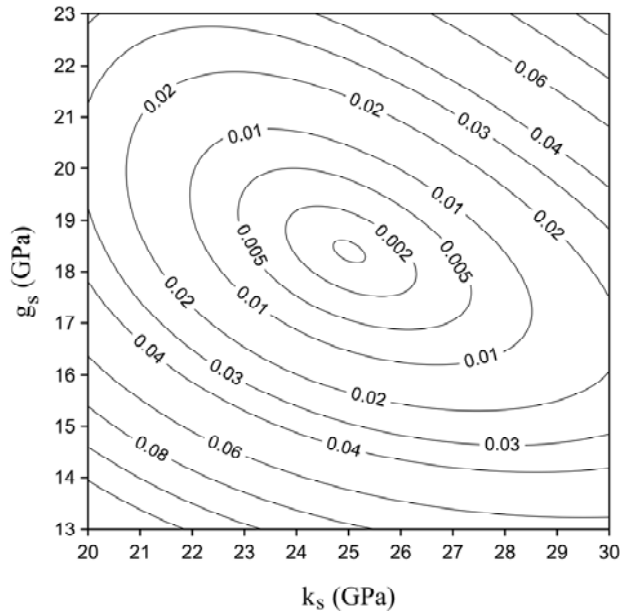

**Figure (2): Contour plot of the calculated error for different values of bulk modulus $k_s$ and shear modulus $g_s$ of the C–S–H solid. The minimum error is equal to $6.1 \times 10^{-4}$, obtained for $k_s$=25.0GPa and $g_s$ =18.4GPa.**

## *5.3. Effect of water-to-cement ratio on poroelastic properties*

The lower w/c limit is chosen equal to 0.4 as this is approximately the lowest w/c ratio for which a complete hydration can be obtained. Figure (3-a) presents the variations of drained and undrained bulk moduli, as well as the variations of Biot and Skempton coefficients with the water-to-cement ratio. We can observe the decrease of the drained and undrained bulk moduli as well as the increase of Biot and Skempton coefficients with w/c ratio.

The decreasing trend of the elastic properties of the hardened cement paste with w/c ratio is compatible with modelling results presented by Sanahuja *et al*. [2] and Pichler et al. [3] and also some experimental results from Helmuth and Turk [62] presented in [2]. Figure (3-b) shows the variations of unjacketed modulus $K_s$ and the modulus $K_\phi$ of the hardened cement paste with w/c ratio. The unjacketed modulus $K_s$ is not significantly influenced by w/c ratio, but this influence is more significant on the modulus $K_\phi$ which shows an increasing trend with w/c ratio. Figure (3-b) shows the calculated active porosity, which increases with w/c ratio. Figure (3-c) shows the significant decrease of shear modulus with w/c ratio. Poisson's ratio also shows a decreasing trend, but only in a narrow range between 0.23 and 0.21.

To verify the model predictions, it is necessary to compare the predicted poroelastic properties with experimentally evaluated parameters. Unfortunately, there are very few experimental measurements of poroelastic properties of the hardened cement paste in the literature except for the dynamic Young's modulus, evaluated from wave velocity or oscillations measurements. The homogenization model is calibrated on static values. One possible solution to compare the model predictions to experimental values from the literature is to use an empirical relation to convert the static modulus calculated by the model into an equivalent dynamic modulus.

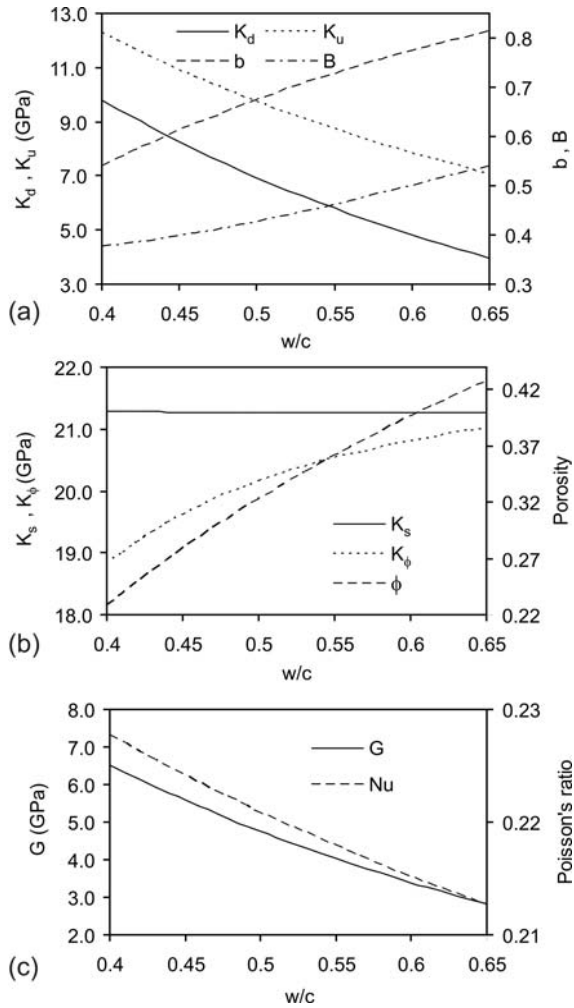

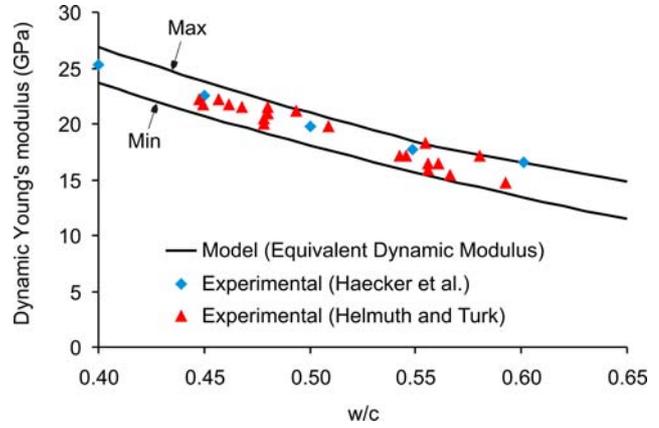

**Figure (4): Comparison of the predicted drained Young's modulus and the experimental results of Helmuth and Turk [62] and Haecker et al. [61]. The static elastic modulus predicted by the model is converted into a range of equivalent dynamic modulus using the empirical relation (61) and the calibration parameters from Hana and Kim [68].**

**Figure (3): Effect of water-to-cement ratio on poroelastic properties**

It is well-known that the dynamic elastic properties of geomaterials, particularly at saturated state, are significantly higher than the static ones [63][64][65][66][67][68]. Different empirical relations between the dynamic and the static elastic moduli of geomaterials are proposed in the literature. Based on the results of an experimental study, Hana and Kim [68] presented empirical relations between the static and dynamic elastic moduli of concrete and showed that curing temperature, age, and cement type do not have a significant influence. They find:

$$E_s = E_d \left(1 - a e^{-b E_d}\right) \qquad (61)$$

where *a* and *b* are calibration parameters. Equation (61) can be solved inversely to calculate the dynamic modulus $E_d$ as a function of the static modulus $E_s$. Hana and Kim [68] presented 8 different sets of parameters *a* and *b* corresponding to samples prepared with different w/c ratio, cement type and curing temperature, but the difference between these calibrations is not very large. The coefficient *a* in different parameter sets varies between 0.402 and 1.021, while the coefficient *b* varies between 0.0170 and 0.0431. The model predictions for the static Young's modulus corresponding to different w/c ratios are converted to equivalent dynamic modulus using equation (61) and all sets of parameters *a* and *b* presented in [68]. The maximum and minimum values of calculated equivalent dynamic moduli are presented in Figure (4) and compared with the experimental results of Helmuth and Turk [62] presented in [2], as well as the experimental results of Haecker *et al*. [61]. The average difference between the maximum and minimum equivalent dynamic modulus is only about 3GPa and we can observe a good compatibility between these limits and the experimentally evaluated dynamic moduli.

# 6. Discussion on the active porosity

For the loading conditions used in our experimental study, a part of the pore volume of the tested hardened cement paste which is most probably situated inside the HD C–S–H is not participating in fluid mass exchange. There is no doubt that the HD C–S–H is a porous phase, but with a lower permeability than the one of the LD C–S–H. The difference between the permeabilities of these phases is probably more highlighted when the cement paste is cured at high temperature, as it is the case for the cement paste used in our experimental study. The microscopic observations of Kjellsen et al. [69] show that the hydration at higher temperatures results in denser inner products. This is compatible with the TEM micrographs of two cement paste hydrated at 20°C and 80°C presented by Richardson [70], which show that the particles of C–S–H formed in the higher temperature are about half the size of those present at lower temperature. Knowing that the HD C–S–H is a porous phase, though with a very low permeability, the activity or the non-activity of its porosity is just a matter of drainage conditions and time. With a slow enough loading rate regarding the actual drainage path, the pore fluid inside the HD CSH should be able to exchange mass with the pore fluid inside the LD C–S–H or the macro-porosity. The porosity of HD C–S–H in this condition should be considered as active porosity.

From an experimental point of view, application of extremely slow loading rates is very difficult, as it increases significantly the duration of the tests and also incorporates much more viscous deformations in the response of the tested sample. This latter effect makes the interpretation and analysis of the test results much more difficult.

The results of our experimental study presented in [4] show that it is hardly possible to establish a perfectly drained condition for the tested hardened cement paste in the time-scale of common laboratory experiments. The loading rate used in drained tests which is determined in a preliminary study is quite slow, equal to 0.025 MPa/min. Even using this loading rate we observed significant effect of viscous strains under confining pressures higher than 40 MPa. Consequently using a lower loading rate for the hardened cement paste, particularly for the experiments under high levels of stress, is not reasonable. Moreover, Biot effective stress analysis of the undrained tests shows a very good compatibility between the results of drained, undrained and unjacketed isotropic compression tests. This very good compatibility which is explained in details in [4], shows that the same drainage conditions have been established in the microstructure of the samples during the tests. In other words, assuming that the inactive porosity is entirely situated inside the HD

C−S−H, the results show that it remained inactive in all three loading conditions: drained, undrained and unjacketed.

The homogenization model enables us to analyse an ideal situation in which the entire porosity of the cement paste is active. This situation corresponds to the application of an extremely slow loading rate, and this analysis enables us to evaluate the poroelastic properties of the hardened cement paste in an ideal, perfectly drained condition which is very difficult to establish in laboratory conditions. In this case the active porosity of the hardened cement paste should be calculated using equation (58) and the volume fractions and porosities presented in Table (4). The calculated active porosity is equal to 0.34 which is very close to the measured total porosity, equal to 0.35. Considering the HD C−S−H as a porous phase imposes some modifications in the equations of the second homogenization step. Consequently, equations (53), (55) and (56) are modified to the following forms:

$$K_{CP}^{hom} = f_{LD} K_{LD}^{hom} A_{LD,CP}^{v} + f_{HD} K_{HD}^{hom} A_{HD,CP}^{v} + f_{CH} k_{CH} A_{CH,CP}^{v} + f_{CK} k_{CK} A_{CK,CP}^{v} \quad (62)$$

$$b^{hom} = 1 - f_{LD} A_{LD,CP}^{v}\left(1 - b_{LD}^{hom}\right) - f_{HD} A_{HD,CP}^{v}\left(1 - b_{HD}^{hom}\right) - f_{CH} A_{CH,CP}^{v} - f_{CK} A_{CK,CP}^{v} \quad (63)$$

$$\frac{1}{N^{hom}} = f_{LD}\left(\frac{\left(1 - A_{LD,CP}^{v}\right)\left(1 - b_{LD}^{hom}\right)}{k_s} + \frac{1}{N_{LD}^{hom}}\right) + f_{HD}\left(\frac{\left(1 - A_{HD,CP}^{v}\right)\left(1 - b_{HD}^{hom}\right)}{k_s} + \frac{1}{N_{HD}^{hom}}\right)$$
$$+ f_{CH}\frac{\left(1 - A_{CH,CP}^{v}\right)}{k_{CH}} + f_{CK}\frac{\left(1 - A_{CK,CP}^{v}\right)}{k_{CK}} \quad (64)$$

The poroelastic parameters of the hardened cement paste are calculated for the ideal, completely drained condition and compared in Table (7) with the results obtained for test conditions, for which the porosity of HD C−S−H is considered as inactive. The comparison shows that the laboratory loading conditions had a very small effect on the results of drained and undrained isotropic compression tests. We observe about 3% decrease in drained bulk modulus, less than 1% decrease in undrained bulk modulus and 5% decrease Skempton coefficient. The most influenced parameter, as expected, is the unjacketed modulus $K_s$ which shows about 25% increase in perfectly drained condition. The unjacketed modulus $K_s$ is an average of the bulk moduli of the different constituents of the microstructure. In test conditions the HD C−S−H behaves undrained and so its undrained bulk modulus, which is considerably smaller than the bulk modulus of C−S−H solid, appears in this average. This results in a smaller unjacketed modulus for test conditions as compared with the perfectly drained condition in which the C−S−H solid bulk modulus appears in the average. The increase of unjacketed modulus $K_s$ results in 15% increase of Biot effective stress coefficient, if evaluated in perfectly drained condition.

| Parameter | $K_d$ (GPa) | $K_s$ (GPa) | $K_u$ (GPa) | $B$ (-) | $G$ (GPa) | $b$ (-) | $\phi_{CP}^{act}$ (-) |
|---|---|---|---|---|---|---|---|
| **Test condition** | 8.5 | 21.3 | 11.2 | 0.39 | 5.8 | 0.60 | 0.27 |
| **Perfectly drained condition** | 8.2 | 26.8 | 11.1 | 0.37 | 5.8 | 0.69 | 0.34 |

**Table (7): Evaluated poroelastic parameters of the tested hardened cement paste for ideal, completely drained condition, comparison with evaluated parameters for test conditions**

# 7. Conclusions

The main purpose of this paper is to extrapolate the results of an experimental study performed on a particular hardened cement paste to other cement pastes corresponding to different water-to-cement ratios. This is done by means of a multi-scale homogenization model which is capable of predicting the poroelastic parameters of a hardened cement paste by knowing the volume fractions and the elastic properties of the constituents of its microstructure. Appropriate models are presented for evaluation of the volume fractions by knowing the chemical composition and density of the cement clinker and water-to-cement ratio. The homogenization model is calibrated using the experimentally evaluated macroscopic poroelastic parameters of the hardened cement paste, which was prepared using a class G cement with w/c=0.44. The model reduces significantly the number of laboratory tests needed to characterize the complete set of the poroelastic parameters of a cement paste. This is a great advantage for experimental studies on hardened cement paste, as due to its very low permeability, the poromechanics tests are usually long.

The association of the results of a macro-scale experimental study with those of micromechanics modelling results in a better understanding of the experimental results and of the phenomena taking place at the micro-scale. Moreover, it permits analysis of the effect of some experimental artefacts on the poroelastic parameters. Based on the results of this study, it seems that in the poromechanics tests the porosity of HD C−S−H was inactive. Considering the very low loading rate in test protocols, this reveals that performing poromechanics tests with a perfect control of the drainage conditions in the microstructure of the hardened cement paste is very difficult. Moreover, it shows that the active porosity of the hardened cement paste for poromechanical calculations should be chosen by taking into account the time-scale and the drainage conditions of the considered problem. The homogenization model permits also to predict the poroelastic parameters in the case of an ideal, perfectly drained condition. This evaluation shows that the testing drainage condition, in which the HD C−S−H porosity is probably inactive, has only a marginal effect, less than 5%, on the drained and undrained poroelastic parameters. It has a greater influence on the unjacketed modulus, causing about 25% underestimation of this modulus in test conditions compared with the perfectly drained condition. The resulting error in the Biot effective stress coefficient is about 15%.

Finally, the capacity of parameter prediction for different conditions and the better understanding of the results of the tests, clearly demonstrate the advantages of the association of macroscopic laboratory experiments and micromechanics modelling.

# 8. Acknowledgment

The author wishes to thank Professor Jean Sulem for fruitful discussions and comments.